\documentclass[12pt,a4paper]{article}

\usepackage{mathrsfs}
\usepackage{amsmath,amsthm,amssymb}
\usepackage[dvips]{graphicx}
\usepackage[abs]{overpic}
\usepackage{wrapfig}
\usepackage{epic,eepic}

\makeatletter

\@addtoreset{equation}{section}
\makeatother

\newcommand{\s}{\hspace{5mm}}

\newcommand{\orddif}[2]{\frac{\mathrm{d} #1}{\mathrm{d} #2}}
\newcommand{\dd}{\mathrm{d}}

\newcommand{\alt}{ \frac{\beta\kappa}{l^2} }
 
\newcommand{\yy}[1]{\beta \big( \mu_#1 + \frac{\nu}{l} \big)}

\newcommand{\Dom}[1]{\textrm{Dom}(#1)}

\def \debugmode{0}

\begin{document}
\vspace*{0.5cm}
\begin{center}
{\Large \bf  Finite Size Effects in Equations of State}\\
\vspace{.3cm}
{\Large \bf under non-trivial Boundary Conditions}
\end{center}
\vspace{0.5cm}
\begin{center}
Nobuhiro Yonezawa
\footnote{
e-mail: yonezawa@post.kek.jp
}
\\
\bigskip
\bigskip
{\em Institute of Particle and Nuclear Studies\\
High Energy Accelerator Research Organization (KEK)\\
Tsukuba 305-0801, Japan}
\end{center}
\bigskip
\begin{abstract}
We study free particles in a one-dimensional box with combinations of two types of boundary conditions:
    the Dirichlet condition and a one-parameter family of quasi-Neumann conditions
        at the two walls.
We calculate energy spectra approximately
    and obtain 
    equations of state having the same (one-dimensional) volume dependence as van der Waals equations of state.
The dependence of the equations of state is examined for particles
    obeying Maxwell-Boltzmann, Bose-Einstein, or Fermi-Dirac statistics. 
Our results suggest that the deviation from ideal gas may also be realized 
    as finite size effects due to
the interaction between the particles and the walls.
\end{abstract}
\newpage
\section{Introduction}
In quantum mechanics,
    boundary conditions play an important role.
A good example is Casimir force \cite{Casimir} predicted in 1948.
It is also known that boundary conditions produce force
    in non-relativistic quantum mechanics \cite{Tsutsui2,Tsutsui3,Tsutsui_2009}.
In this paper, we focus on one-dimensional free particles
    and show that this force leads to
    equations of state whose
        one-dimensional volume or length
            dependence is the same as that of van der Waals equation of state.
\par
Consider a closed box.
One may impose the conventional Dirichlet condition on wave functions at the walls of the box.
In mathematics,
    we are, however, permitted to generalize the boundary condition
        in the context of self-adjoint extensions of Hamiltonian operators \cite{Read_Simon},
            where the Dirichlet and Neumann conditions are allowed just
                as special boundary conditions.
Interestingly, it has been pointed out \cite{Tsutsui6} that the generalized boundary conditions 
    can be derived from the vanishing limits of widths of two step functions,
    which may be realized at, for example, junctions of two semiconductor layers.                      
In this paper, we consider the Dirichlet condition and a class of boundary conditions including the Neumann condition,
    which we call quasi-Neumann conditions,
        to find how boundary conditions affect the physical property of the particles in the box.
We show that such boundary conditions affect  
    statistical quantities of finite size systems. Note that this effect  
        has no relation with bulk properties of the system in the  
            thermodynamic limit, where the effect that we will obtain disappears.
\par
Specifically, we approximately calculate the spectra of particles in the box
    by integral approximation
    and find that the spectra can lead to 
    equations of state
    having the same dependence on length as van der Waals equation of state.
Van der Waals equation was
    introduced by Johannes Diderik van der Waals
    in his Ph.D. thesis \cite{v_d_Waals} and has been
        well-known to describe the behavior of real fluids.
Van der Waals-like equations of state were derived from, for example, 
    the Gaussian Markoffian process \cite{Kac},
    the intermolecular interaction in classical statistical mechanics \cite{Landau_Lifshitz,Chan},
    the wall-molecular interaction in classical statistical mechanics \cite{Sullivan},
    or the electro-magnetic field \cite{Lifshitz}.
Unlike van der Waals equation of state,
        our equations of state tend to that for ideal gas in the thermodynamic limit.
	       In finite size system,
    the quantum boundary effects lead to similar equations of state as the above-mentioned effects.  
The free particles under the Dirichlet condition \cite{Grossmann} and cyclic boundary condition \cite{Toms} have been studied earlier,
    and here we consider combinations of the non-trivial boundary conditions
    and show that they admit
equations of state
whose length dependence is the same as that of van der Waals equation of state.  
\par
\begin{figure}[htpb]
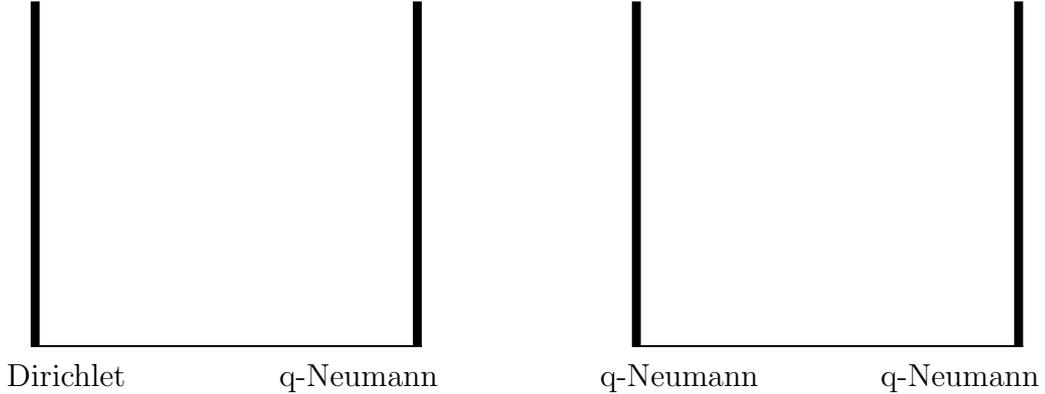
      
    \begin{center}

    \end{center}
    \caption{{\footnotesize
        The energy eigenfunctions and the energy levels in the boxes with different boundary conditions.
        The solid lines show the eigenfunctions.
        The height of dashed lines indicates the square root of the energy levels.
        }}
    \label{fig:boxes}
\end{figure}

This paper is organized as follows.
In section 2,
    we briefly review the method of self-adjoint extensions to characterize the possible boundary conditions.
In section 3,
    we study free particles in a one-dimensional box
        with
            the Dirichlet condition at its one wall
            and
            a quasi-Neumann condition at the other wall (see the left one of Fig.\ref{fig:boxes}).
We consider three types of statistics:
    Maxwell-Boltzmann statistics,
    Bose-Einstein statistics,
    and
    Fermi-Dirac statistics,
        for which equations of state are obtained.
In section 4,
    we study
\ifodd \debugmode
\textbf{free particles in}
\else
    free particles in
\fi
a one-dimensional box with identical boundary conditions
        at its both walls (see the right one of Fig.\ref{fig:boxes}),
            where we assume Maxwell-Boltzmann statistics.
In addition,
    we extend the one-dimensional box with identical quasi-Neumann conditions at both of the walls to a three-dimensional one.
In section 5,
    we summarize our results and discuss the physical meanings on the outcomes.
\section{On self-adjoint extensions of the Hamiltonian}
In this section, we give a short review on self-adjoint extensions based on \cite{Tsutsui4}. 
Consider particles in a one-dimensional box with the length $l$.
Each particle is governed by the Hamiltonian,
\begin{align}
H=&-\frac{\hbar^2}{2m}\orddif{^2\  }{x^2},
\label{eq:Hamiltonian}
\end{align}
where $x \in [a,b]$ is a coordinate in the box.   
The Dirichlet conditions may be imposed on a wave function at the walls of the box.
In general, infinitely large number of pairs of boundary conditions at both of the walls is permitted
    as long as the Hamiltonian is self-adjoint in mathematics.
It is known that each pair of boundary conditions defines a distinctive wave function and spectrum.
Thus, this generalization of boundary conditions has physical meaning.
\par
Let $\Dom{H}$ be the domain of the Hamiltonian.
If $H$ is a self-adjoint operator,
    then $\phi_1, \phi_2\in \Dom{H}$ must satisfy
\begin{align}
0 &
= \int ^b_a \dd x \ 
        \phi_1^*(x) \Big\{ 
            -\frac{\hbar^2}{2m}\orddif{^2}{x^2}\phi_2(x) 
        \Big\}
    - \int ^b_a \dd x 
        \Big\{-\frac{\hbar^2}{2m}\orddif{^2}{x^2}\phi_1(x)
    \Big\}^* \phi_2(x)\notag\\
&=-\frac{\hbar^2}{2m}
    \Big\{
        \phi_1^*(b) \orddif{\phi_2}{x}(b)
        -\orddif{\phi_1^*}{x}(b) \phi_2(b)
    \Big\}
+\frac{\hbar^2}{2m}
    \Big\{
        \phi_1^*(a) \orddif{\phi_2}{x}(a)
        -\orddif{\phi_1^*}{x}(a) \phi_2(a)
    \Big\}.
\label{eq:self-adjoint1}
\end{align}
Note that 
    the above equations represent
        the conservation of the probability current density at both of the walls
            if $\phi_1=\phi_2$.
We assume that the wall of $x=a$ is disconnected from that of $x=b$,
and for example,
\ifodd \debugmode
    \textbf{the cyclic boundary condition is excluded.}
\else
            the cyclic boundary condition is excluded.
\fi
\ifodd \debugmode
    \textbf{The disconnected condition implies }
\else
            The disconnected condition implies
\fi
that each term of the second line of (\ref{eq:self-adjoint1}) 
    must be equal to $0$ independently.
\ifodd \debugmode
\textbf{We discuss only the boundary at $x=a$,
since the boundary condition at $x=b$ can be dealt with similarly.}
\else
        We discuss only the boundary at $x=a$,
since the boundary condition at $x=b$ can be dealt with similarly.
\fi
\par
For convenience,
    we use $\phi^\pm_i (x)$ introduced as
\begin{align}
\phi^\pm_i (x) := \phi_i(x) \pm i L \orddif{\phi_i}{x} (x),
\end{align}
where $L \in \mathbb{R}$ and $i = 1, 2$.
Note that $L$ has the dimension of length.
We can rewrite the condition (\ref{eq:self-adjoint1}) as
\begin{align}
{\phi_1^+}^*(a) \phi_2^+(a) = {\phi_1 ^-}^*(a) \phi_2^-(a).
\end{align}
The above equation means that
    the (one-dimensional) Hermite inner product of $\phi_1^+(a)$ and $\phi_2^+(a)$
    is equal to that of $\phi_1^-(a)$ and $\phi_2^-(a)$.
Since $\phi_1$ and $\phi_2$ are arbitrary elements of $\Dom{H}$,
    $\phi^+_i$ must be connected with $\phi^-_i$
        by a (one-dimensional) unitary transformation specified by $\Dom{H}$:
\begin{align}
e^{i \theta} \phi^+_i (a) &= \phi^-_i (a).
\end{align}
Note that $\Dom{H}$ is characterized by the unitary transformation vice versa.
This procedure brings the condition of self-adjointness to a simple form:
\begin{align}
(e^{ i \theta} - 1) \phi(a)+ i L (e^{ i \theta} + 1) \orddif{\phi}{x}(a) &= 0.
\label{eq:self-adjoint2}
\end{align}
This equation is known as Cheon-F\"{u}l\"{o}p-Tsutsui boundary equation\cite{Tsutsui1}.
\par
We introduce $L_\theta := - L \cot \frac{\theta}{2}$ instead of $L$ and $\theta$,
then (\ref{eq:self-adjoint2}) can be written as follows:
\begin{align}
 \phi(l) = L_\theta \orddif{\phi}{x}(l).
 \label{eq:self-adjoint3}
\end{align}
The above equation contains the arbitrary parameter $L_\theta$.
In mathematics,
    the Hamiltonian (\ref{eq:Hamiltonian}) is not a bounded operator;
        therefore it
\ifodd \debugmode
    \textbf{cannot be defined in the entire}
\else
    cannot be defined in the entire 
\fi
    Hilbert space but a dense subspace of it.
The domain of the Hamiltonian cannot be specified uniquely, either.
The arbitrary parameter $L_\theta$ implies this ununiqueness.
In particular,
    $L_\theta = 0$ and $L_\theta = \infty$ imply the Dirichlet condition and the Neumann condition, respectively.
\ifodd \debugmode \textbf{
Note that the boundary condition (\ref{eq:self-adjoint3}) has been derived \cite{Tsutsui6}
    from the limits of two step functions,
    whose ratio between the heights and widths characterizes $L_\theta$.
}
\else
Note that the boundary condition (\ref{eq:self-adjoint3}) has been derived \cite{Tsutsui6}
    from the limits of two step functions,
    whose ratio between the heights and widths characterizes $L_\theta$   
\fi
\section{The box with the Dirichlet and a quasi-Neumann conditions}
\label{sec:D_N}
In this section, we study the box with
    the Dirichlet
    and
    a quasi-Neumann conditions.
\par 
Let $x \in [0,l]$ be a coordinate in the box.
We assume that our wave function satisfies
    the Dirichlet condition at $x=0$
    and
    a boundary condition characterized by $L_\theta$ at $x=l$.
We restrict
\ifodd \debugmode
    \textbf{ourselves to $|L_\theta| > l$ at $x=l$, which is required technically to ensure the validity of our approximation.}
\else
            ourselves to $|L_\theta| > l$ at $x=l$, which is required technically to ensure the validity of our approximation.
\fi
A class of such boundary conditions includes
        the Neumann condition,
        but not
        the Dirichlet condition;
\ifodd \debugmode
    \textbf{accordingly,}
\else
            accordingly,
\fi
we call such a boundary condition a quasi-Neumann condition.
\par
A wave function of each particle is given by
\begin{align}
    \phi(x)&=A \sin kx, \s (k \in \mathbb{R}),
        \label{eq:positive eigenvalue case}
\end{align}
and its spectrum condition is obtained by
\begin{align}
\frac{l}{L_\theta} = kl \cot kl.
    \label{eq:quantization condition}
\end{align}
Note that the above equation is a transcendental equation;
    therefore we cannot solve it explicitly.
\subsection{Calculation of approximated spectrum}
\begin{figure}[t]
  \begin{center}  
    \includegraphics{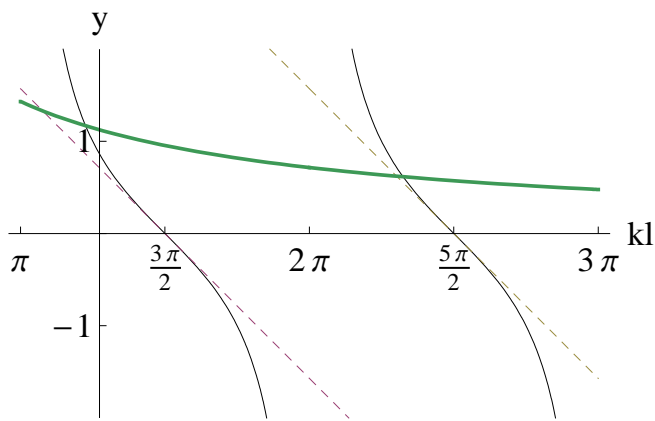}
  \end{center}
  \caption{{\footnotesize
    The thin lines represent $y = \cot kl$ and
    the dashed lines show their linear approximation. 
    The thick line is $y=\frac{l}{L_\theta}\frac{1}{kl}$.
    The crossing points of the thick and thin lines yield the solutions of (\ref{eq:quantization condition}).
    Those of the thick and dashed lines yield the approximation solutions.
  }}
  \label{fig:explanation_of_linear_approximation}
\end{figure}

\ifodd \debugmode
\textbf{Now we study the equation (\ref{eq:quantization condition})
    and obtain its approximated solutions.}
\else
        Now we study the equation (\ref{eq:quantization condition})
    and obtain its approximated solutions.
\fi
\ifodd \debugmode \textbf{The spectrum condition (\ref{eq:quantization condition}) has a solution 
}\else
The spectrum condition (\ref{eq:quantization condition}) has a solution 
\fi for each region,
    $(n-1)\pi < kl <n\pi$,
    where $n \in \mathbb{N}$.
In addition,
\ifodd \debugmode
\textbf{the solution}
\else the solution \fi
 $kl$ tends to $\pi(n-\frac{1}{2})$ in the limit $n \to \infty$ (see Fig.\ref{fig:explanation_of_linear_approximation}).
We expand $\cot kl $ as 
\begin{align}
\cot kl = \Big( n - \frac{1}{2} \Big) \pi - kl + O \Big[ \Big(kl - n \pi + \frac{1}{2}\pi \Big) ^3 \Big].
\label{eq:aproximation1}
\end{align}
\par
For each region, the following inequality is fulfilled:
\begin{align}
\Big| \Big( n-\frac{1}{2} \Big) \pi -kl \Big| < \big| \cot kl \big| = \Big|\frac{l}{L_\theta}\frac{1}{kl} \Big|.
\end{align}
This inequality leads to 
\begin{align}
\Big|(n-\frac{1}{2})\pi -kl \Big| < \Big| \frac{l}{L_\theta}\frac{1}{\pi \big( n-\frac{1}{2} \big) }\Big| 
+ O\Bigg[ \bigg\{ 
    \frac{l}{L_\theta}\frac{1}{\pi \big( n-\frac{1}{2} \big) } 
\bigg\}^2 \Bigg].
\label{eq:error_estimation}
\end{align}
The approximation solution of (\ref{eq:quantization condition})
    for each region, $(n-1)\pi < kl <n\pi$,
is therefore obtained by
\begin{align}
kl
&=
\pi \Big( n-\frac{1}{2} \Big) - \frac{l}{\pi L_\theta \big( n -\frac{1}{2} \big)}
+ O\Bigg[ \bigg\{ 
    \frac{l}{L_\theta}\frac{1}{\pi \big( n-\frac{1}{2} \big) } 
\bigg\}^3 \Bigg],
\end{align}
where the error term is estimated from (\ref{eq:error_estimation}).
The solutions lead to energy spectrum of each particle $E_n$ as\footnote{
    The approximated spectrum (\ref{eq:Energy_of_N_D}) and (\ref{eq:Energy_of_N_N}) were studied
        in a more general context: e.g., \cite{Levitan},
        where the approximation is valid in the region $n \ge n_0$ for some $n_0$.
    Our formulae (\ref{eq:Energy_of_N_D}) and (\ref{eq:Energy_of_N_N}) are valid
        for all $n \in \mathbb{N}$.
}
\begin{align}
E_n &=
    \frac{\kappa}{l^2}
        \bigg\{
            1
            +
            \Big(
                \frac{l}{L_\theta}
            \Big)^2 
            \frac{1}{ \pi^4 \big( n-\frac{1}{2} \big)^4}
        \bigg\}
    \Big( n - \frac{1}{2} \Big)^2
    - \frac{\nu}{l}\Bigg[
    1-
    O \bigg[
        \bigg\{ 
            \frac{l}{L_\theta} \frac{1}{ \pi \big( n-\frac{1}{2} \big)}
        \bigg\}^2
    \bigg]
    \Bigg]\notag\\
&\simeq 
    \frac{\kappa \big( n - \frac{1}{2} \big)^2}{l^2}
    - \frac{\nu}{l},
\label{eq:Energy_of_N_D}
\end{align}
where
\begin{align}
\kappa&:=\frac{\hbar^2\pi^2}{2m},&
\nu&:=\frac{\hbar^2}{m L_\theta},
\end{align}
and we ignore the terms of the order $\frac{l}{L_\theta}$ 
\ifodd \debugmode
    \textbf{by our quasi-Neumann condition.}
\else
            by our quasi-Neumann condition.
\fi
\subsection{Statistical quantity and equation of state}
In this section,
    we calculate statistical quantities and derive equations of state from the spectrum
    in high temperature region $\frac{\beta\kappa}{l^2}<1$.
We assume $N$ particles obeying 
    Maxwell-Boltzmann statistics,
    Bose-Einstein statistics,
    and
    Fermi-Dirac statistics.
\subsubsection{The case of Maxwell-Boltzmann statistics}
\label{sec:The_case_of_Maxwell-Boltzmann_statistics}
Now, we study the case of Maxwell-Boltzmann statistics.
By using (\ref{eq:integral_approximation_of_one_half}),
    we obtain partition function $Z_c$ as follows:
\begin{align}
Z_c 
=& \bigg\{ \sum_{n=1}^\infty \exp( -\beta E_n ) \bigg\}^N \notag\\
=&
    \exp \Big( \frac{N \beta \nu}{l} \Big)
        \bigg\{ \int^\infty_0 \dd x
            \exp \Big( -\frac{\beta \kappa}{l^2} x^2 \Big)
        \bigg\}^N \notag\\
=&
    \Big( \frac{ l }{2 }
        \sqrt{\frac{ \pi}{ \beta \kappa }}
    \Big)^N
            \exp \Big(\frac{N\beta \nu}{l}\Big).
\end{align}
Let $p_c$ be force acting on the walls of the box.
We can, based on statistical mechanics, derive $p_c$ from following equation:
\begin{align}
p_c= \frac{1}{\beta Z_c}\orddif{Z_c}{l}.
\end{align}
The equation of state in this case is thus given by
\begin{align}
\Big(p_c +\frac{N \nu}{l^2}\Big) l
&=
    \frac{N}{\beta}.
\label{eq:classical_equation_of_state_at_N-D}
\end{align}
The given equation differs from 
\ifodd \debugmode
    \textbf{ideal gas law}
\else
            ideal gas law
\fi
in the term $\frac{N \nu}{l^2}$.
\ifodd \debugmode
    \textbf{We refer to this term as}
\else
            We refer to this term as
\fi
a force correction term.
Note that (\ref{eq:classical_equation_of_state_at_N-D}) becomes
\ifodd \debugmode
    \textbf{the equation of state for ideal gas}
\else
            the equation of state for ideal gas
\fi
if the wave function satisfies the Neumann condition at $x = l$.
\ifodd \debugmode \textbf{This implies that
    equation of state for ideal gas cannot be derived under a pair of Dirichlet conditions,
    which will be discussed in section \ref{subsec:D-D}.
}\else This implies that
    equation of state for ideal gas cannot be derived under a pair of Dirichlet conditions,
    which will be discussed in section \ref{subsec:D-D}.
\fi
\par
The equation of state (\ref{eq:classical_equation_of_state_at_N-D}) resembles van der Waals equation of state:
\begin{align}
\Big( p_{\textrm{vdW}} + \frac{N^2 \nu_{\textrm{vdW}}}{l^2} \Big)(l- \sigma_{\textrm{vdW}} N) = \frac{N}{\beta}.
\label{eq:vdW_equation_of_state}
\end{align}
Van der Waals equation of state is different from ideal gas law
    in the terms $\frac{N^2 \nu_{\textrm{vdW}}}{l^2}$ and $- \sigma_{\textrm{vdW}} N$. 
We call
    the former a force correction term
    and
    the latter a length correction term.
Van der Waals equation differs
    from (\ref{eq:classical_equation_of_state_at_N-D}) in two points.
\ifodd \debugmode
\textbf{One is that}
\else
        One is that
\fi
van der Waals equation of state has the length correction term,
    while (\ref{eq:classical_equation_of_state_at_N-D}) does not.
In section \ref{sec:other_boundary_conditions},
    we will derive equations that possess length correction terms, which are however different from van der Waals one.
\ifodd \debugmode \textbf{The other is that
}\else The other is that
\fi
the force correction term in (\ref{eq:classical_equation_of_state_at_N-D}) is in proportion to $N$,
\ifodd \debugmode
    \textbf{while}
\else
            while
\fi
that of van der Waals equation of state is in proportion to $N^2$;
\ifodd \debugmode
    \textbf{therefore,
the force correction term in (\ref{eq:classical_equation_of_state_at_N-D}) vanishes in the thermodynamic limit $l\to\infty$ under the constant density.
This is reasonable because the boundary effects are expected to disappear in the thermodynamic limit
    where the property of bulk gas becomes dominant.
Our equation is thus meaningful only in finite length systems.}
\else
            therefore,
the force correction term in (\ref{eq:classical_equation_of_state_at_N-D}) vanishes in the thermodynamic limit $l\to\infty$ under the constant density.
This is reasonable because the boundary effects are expected to disappear in the thermodynamic limit,
    where the property of bulk gas becomes dominant.
Our equation is thus meaningful only in finite length systems.
\fi
We will further discuss this point in the last section.
\subsubsection{The cases of Bose-Einstein statistics and Fermi-Dirac statistics}
Now, we study the case of quantum statistics.
Let bosonic variables be subscripted by $+$
    and
    fermionic variables by $-$.
\par
\ifodd \debugmode \textbf{We
}\else We
\fi
express, by using integral approximation (\ref{eq:integral_approximation_of_one_half}),
        the particle number $N$
        as a function of chemical potential $\mu_\pm$ like that:
\begin{align}
N
=&
    \sum^\infty_{n=1}
        \frac{1}{
            \exp \big\{ \beta(E_n-\mu_\pm) \big\} \mp 1
        }\notag\\
=& \pm \frac{l}{2}\sqrt{\frac{\pi}{\beta\kappa}} 
    \textrm{Li}_\frac{1}{2} \Big[
        \pm \exp \Big \{ \yy{\pm} \Big\}
    \Big],
\label{eq:num_vs_chemical_p}
\end{align}
where $\textrm{Li}_s (x)$ is polylogarithm function
    (see Appendix \ref{sec:polylog}).
Note that the integral approximation of the bosonic case is violated 
    in the region $\yy{+}\sim 0$
since the integrand becomes infinity at $\yy{+} = 0$.
We will discuss this case in section \ref{subsec:limitation}.
\par
\ifodd \debugmode
\textbf{Analogously,
the force in this case $p_{\pm}$ is obtained by}
\else
Analogously,
the force in this case $p_{\pm}$ is obtained by
\fi
\begin{align}
p_{\pm}
=&-\sum^\infty_{n=1}\orddif{E_n}{l}
        \frac{1}{
            \exp \{ 
                \beta(E_n-\mu_-)
            \} 
            \mp 1
        }
\notag\\
=&\sum^\infty_{n=1}
        \frac{
            \frac{2 \kappa}{l^3} \big( n-\frac{1}{2} \big) ^2
            -\frac{\nu}{l^2}
        }{
            \exp[\beta(E_n-\mu_-)]\mp 1
        }\notag\\
=&
    \pm\frac{1}{2 \beta} 
        \sqrt{\frac{\pi}{\kappa \beta}}
            \textrm{Li}_\frac{3}{2} \Big[
                \pm \exp \Big\{ \beta(\frac{\nu}{l}+\mu_\pm) \Big\}
            \Big]
        -\frac{N \nu}{l^2}.
\label{eq:quantum_equation_of_state_at_D-N_0} 
\end{align}
\par
\begin{figure}[t]
 \begin{minipage}{0.5\hsize}
  \begin{center}
    \includegraphics[keepaspectratio=true,height=50mm]{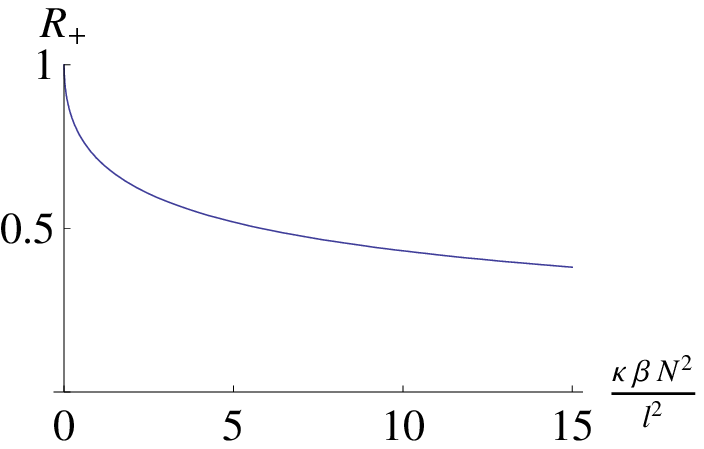}
  \end{center}
   \end{minipage}
 \begin{minipage}{0.5\hsize}
  \begin{center}
    \includegraphics[keepaspectratio=true,height=50mm]{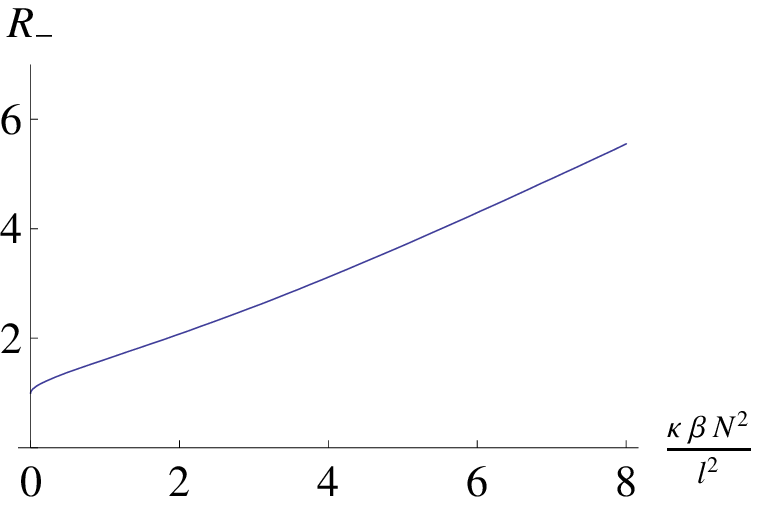}
  \end{center}
     \end{minipage}
  \caption{\footnotesize{
    These figures show the behavior of $R_\pm$.
\ifodd \debugmode
}\textbf{(Left) Bosons.}
\footnotesize{\else
    (Left) Bosons.
\fi
    It shows that $R_+$ is a monotonically decreasing function of $\frac{\kappa \beta N^2}{l^2}$.
\ifodd \debugmode
}\textbf{(Right) Fermions.}
\footnotesize{\else
    (Right) Fermions.
\fi
    It shows that $R_-$ is a monotonically increasing function of $\frac{\kappa \beta N^2}{l^2}$.
    The graphs imply that both $R_\pm$ tend to $1$ in the limit $\frac{\kappa \beta N^2}{l^2} \to 0$.
  \label{fig:x_vs_R}
  }}
\end{figure}

We can rewrite the equation (\ref{eq:num_vs_chemical_p}) as
\begin{align}
\frac{\kappa\beta N^2}{l^2} 
= \frac{\pi}{4} \bigg[ 
    \textrm{Li}_\frac{1}{2} \big\{
        \pm e^{ \beta(\mu_\pm+\frac{\nu}{l}) }
    \big\} \bigg]^2.
\end{align}
Since $\textrm{Li}_\frac{1}{2}(\pm e^y) $ is a monotone function of $y$,
    there 
\ifodd \debugmode
    \textbf{exists}
\else
            exists			
\fi  
     the inverse function; 
\ifodd \debugmode \textbf{
therefore,
}\else
therefore,
\fi
$\beta \big( \mu_+ + \frac{\nu}{l}\big)$ can be regarded as a function of $\frac{\kappa \beta N^2}{l^2}$.
\ifodd \debugmode
\textbf{For this reason, we introduce $R_\pm \Big( \frac{\kappa \beta N^2}{l^2} \Big)$ by}
\else
For this reasons, we introduce $R_\pm \Big( \frac{\kappa \beta N^2}{l^2} \Big)$ by
\fi
\begin{align}
R_\pm \Big( \frac{\kappa \beta N^2}{l^2} \Big):=
    \frac{\textrm{Li}_\frac{3}{2}
            \Big[
                \pm \exp \Big\{ \beta(\frac{\nu}{l}+\mu_\pm) \Big\}
            \Big]
    }{
        \textrm{Li}_{\frac{1}{2}}
            \Big[
                \pm \exp \Big\{ \beta(\frac{\nu}{l}+\mu_\pm) \Big\}
            \Big]
    },
\end{align}
whose behavior is shown in Fig. \ref{fig:x_vs_R}.
\par
From the above equation, we can rewrite (\ref{eq:quantum_equation_of_state_at_D-N_0}) as
\begin{align}
\Big( p_\pm + \frac{N\nu}{l^2} \Big) l 
= \frac{N}{\beta}
    R_\pm \Big( \frac{\kappa \beta N^2}{l^2} \Big).
\label{eq:quantum_equation_of_state_at_D-N}
\end{align}
\ifodd \debugmode \textbf{
We regard the above equation as the equation of state in this case, which 
}\else
We regard the above equation as the equation of state in this case, which
\fi
also has the term, $\frac{N\nu}{l^2}$, as is the case for (\ref{eq:classical_equation_of_state_at_N-D}).
We
\ifodd \debugmode \textbf{
call this term
}\else
call this term
\fi
a force correction term as well. 
In the bosonic case, 
    a similar equation without the force correction term was derived from the cyclic boundary condition in \cite{Toms}.
Note that the equation (\ref{eq:classical_equation_of_state_at_N-D}) is derived from replacing $R_\pm$ with $1$.
\subsection{Behavior in the limit $l \to 0$}
\label{subsec:limitation}
In the previous section, we calculate the statistics quantities in the high temperature region $\frac{\beta \kappa}{l^2} < 1$.
In this section,
    we study the behavior of the equations of state in the limit $\beta \to \infty$.
\subsubsection{The case of Maxwell-Boltzmann statistics}
\label{sec:limitation_attitude_of_classical_particles}
In the region $\frac{\kappa\beta}{l^2} \gg 1$,
the integral approximation is not better than first term approximation:
\begin{align}
Z_c = \exp (-\frac{\beta N \kappa}{4 l^2}+\frac{\beta N \nu}{l}).
\end{align}
Then $p_c$ is given by
\begin{align}
\big( p_c + \frac{N \nu }{l^2} \big) \ l = \frac{\kappa N }{2 l^2}.
\label{eq:lim_of_classical_equation_of_state}
\end{align}
We call the term $\frac{N \nu}{l^2}$ a force correction term as well.
Note that the term is the same shape
    as those of (\ref{eq:classical_equation_of_state_at_N-D}) and (\ref{eq:quantum_equation_of_state_at_D-N}).
\subsubsection{The case of Bose-Einstein statistics}
In the bosonic case,
    the integral approximation is not verified in the region $\yy{+} \sim 0$,
\ifodd \debugmode
\textbf{since}
\else
since
\fi
the integrand of (\ref{eq:integral_approximation_of_one_half}) diverges at $\yy{+} = 0$;
therefore the second term of the right hand side of (\ref{eq:integral_approximation_of_one_half})
\ifodd \debugmode
\textbf{cannot be neglected}. 
\else
cannot be neglected. 
\fi
In this region, we use the following approximation:
\begin{align}
N&=\sum_{n=1}^\infty
    \frac{1}{
        \exp \big\{ \alt \big( n-\frac{1}{2} \big)^2 - \yy{+} \big\}-1
    }
\notag\\
&   
=\frac{1}{
    \exp \Big\{ \frac{\beta\kappa}{4 l^2} - \yy{+} \Big\}-1
} 
+
\int_1^\infty
    \frac{\dd x}{
        \exp \Big\{ \alt x^2 - \yy{+} \Big\}-1
    }.
\label{eq:BEC_reg}
\end{align}
An upper bound of the integral term is given by
\begin{align}
    \int_1^\infty
        \frac{dx}{
            \exp \big\{ \alt x^2 - \yy{+} \big\}-1
        }
<&
    \int^\infty_1
        \frac{dx}{
            \alt x^2 - \yy{+} -1
        }\notag\\
<&
    \frac{l^2}{\beta\kappa}.
\end{align}
\par
If $\frac{N}{2} \gg  \frac{l^2}{\beta\kappa}$,
    the first term of the second line of (\ref{eq:BEC_reg}) is much bigger than the integral term.
In other words,
    the average number of the particles in the ground state
        is much bigger than 
\ifodd \debugmode \textbf{that
}\else that
\fi
in the excited states.
Bose-Einstein condensation therefore happens in that region.
Note
    that the condition, $\frac{N}{2} \gg  \frac{l^2}{\beta\kappa}$, is independent of the boundary parameter $L_\theta$. 
In this region,
    we can neglect
the integral term of (\ref{eq:BEC_reg});
    therefore the equation of state is given by as follows:
\begin{align}
\big( p_+ + \frac{N \nu }{l^2} \big) \ l = \frac{\kappa N }{2 l^2}.
\label{eq:lim_of_bose_equation_of_state}
\end{align}
The above equation is identical to (\ref{eq:lim_of_classical_equation_of_state})
    based on Maxwell-Boltzmann statistics. 
\subsubsection{The case of Fermi-Dirac statistics}
In the fermionic case, the equation of state (\ref{eq:quantum_equation_of_state_at_D-N}) is still
    valid in the limit of $\beta \to \infty$.
From appendix \ref{sec:polylog},
the asymptotic behavior of $R_-(z)$ in the limit of $z \to \infty$ is 
\begin{align}
R_-(z) \to &\frac{2}{3} z.
\end{align}
The equation of state thus tends to
\begin{align}
\big( p_- + \frac{N \nu}{l^2} \big) \ l = \frac{2 \kappa N^3}{3 l^2}.
\label{eq:lim_of_fermi_equation_of_state}
\end{align}
We also call the term $\frac{N \nu}{l^2}$ a force correction term.
Note that the force correction term is the same shape
    as those of
        (\ref{eq:classical_equation_of_state_at_N-D}),
        (\ref{eq:quantum_equation_of_state_at_D-N}),
        (\ref{eq:lim_of_classical_equation_of_state}),
        and
        (\ref{eq:BEC_reg}).
\section{Pairs of identical boundary conditions}
\label{sec:other_boundary_conditions}
We have studied 
    the statistical quantities of the system in
        the box with the Dirichlet condition and a quasi-Neumann condition
            in the previous section.
In this section, we assume pairs of identical boundary conditions:
    a pair of the Dirichlet conditions
    and
    a pair of identical quasi-Neumann conditions.
In addition, we extend the one-dimensional box with a pair of quasi-Neumann conditions
    to a three-dimensional box.
We consider
    particles obeying Maxwell-Boltzmann statistics 
    since the quantum statistics cases are hard to study by using the method
    developed in section \ref{sec:D_N}.\footnote{
The particle number is not a monotone function like (\ref{eq:num_vs_chemical_p});
    therefore, we cannot repeat the discussion of section \ref{sec:D_N}.
}
\subsection{One-dimensional box with the Dirichlet condition}
\label{subsec:D-D}
In section \ref{sec:The_case_of_Maxwell-Boltzmann_statistics},
    we have shown that
        the equation of state
    under Maxwell-Boltzmann statistics with       
        the pair of
            the Dirichlet condition
            and
            the Neumann condition
    is identical to        
    that for ideal gas.
In this section, we show that the equation of state of the pair of the Dirichlet conditions is
    different from that for ideal gas.
\par
We impose the Dirichlet condition at the walls of the box.
Let $E_n$ be energy of each particle given by
\begin{align}
E_n=\frac{\kappa n^2}{l^2},
\end{align}
where $n\in\mathbb{N}$.
By using (\ref{eq:integral_approximation_of_integer}),
    we obtain partition function $Z^{DD}$ as follows:
\begin{align}
Z^{DD} &= \big\{ \sum_{n=1}^\infty \exp ( -\beta E_n) \big\}^N \notag\\
&= \Big\{ \int^\infty_0 \dd x \exp \big( -\frac{\beta\kappa}{l^2} x^2 \big) - \frac{1}{2}\Big\}^N \notag\\
&= \Big( \frac{l}{2}\sqrt{\frac{\pi}{\kappa \beta}} -\frac{1}{2} \Big)^N.
\end{align}
Let $p^{DD}$ be the force acting on the wall.
The equation of state in this case is thus given by
\begin{align}
p^{DD} \ 
    \bigg( l - \sqrt{\frac{\beta\kappa}{\pi}} \bigg)
=
    \frac{N}{\beta}.
\label{eq:classical_equation_of_state_at_D-D}
\end{align}
We call the term $- \sqrt{\frac{\beta\kappa}{\pi}}$ a length correction term.
The above equation of state is clearly different from 
        ideal gas law,
which is derived from classical statistical mechanics;
    therefore quantum effect is responsible for this difference.
\par
Tsutsui \textit{et al.} have studied a one-dimensional box with a partition
    at the center of the box \cite{Tsutsui2,Tsutsui3}.
They imposed the Dirichlet conditions on wave functions at the walls of the box.
They also imposed the Dirichlet condition at one side of the partition
    and the Neumann condition at the other side of the partition.
It was shown 
that the partition is, in high temperature limit $\beta \to 0$, subjected to force $\Delta p$ given by
\begin{align}
\Delta p = \frac{N}{l^2} \sqrt{\frac{\kappa}{\beta\pi}},
\label{eq:tsutsui_pressure}
\end{align}
where $l$ is the length of each half of the box. 
\par
Below, we derive $\Delta p$ from the equations of state (\ref{eq:classical_equation_of_state_at_N-D}) and (\ref{eq:classical_equation_of_state_at_D-D}).
In the high temperature limit,
    the length correction term is smaller than $l$;
therefore we approximately rewrite $p^{DD}$ as follows:
\begin{align}
p^{DD}
&=\frac{N}{l \beta}
    \frac{1}{
        1-\frac{1}{l}\sqrt{\frac{\beta\kappa}{\pi}}
    }\notag\\
&\sim
\frac{N}{l \beta}
+\frac{N}{l^2} \sqrt{\frac{\kappa}{\beta\pi}}\notag\\
&=
p^{DN}+ \Delta p,
\end{align}
where $p^{DN}$ is the force of Maxwell-Boltzmann particles in the box
    with a pair of the Dirichlet condition and the Neumann condition.
Since the difference of the forces $p^{DD}-p^{DN}$ is given by $\Delta p$,
our results agree with that of \cite{Tsutsui2,Tsutsui3} in the high temperature limit.
\subsection{One-dimensional box with a quasi-Neumann Condition}
\label{subsec:CN-CN}
In this section, we study particles in the box with a quasi-Neumann condition.
Let $x \in [-\frac{l}{2},\frac{l}{2}]$ be a coordinate in the box.
We assume identical boundary conditions at the walls.
In accordance with \cite{Tsutsui5},
    the boundary conditions at $x= \frac{l}{2}$ and $x=-\frac{l}{2}$ are characterized by the boundary parameters $L_\theta$ and $-L_\theta$, respectively.
The sign change of the boundary parameter 
is caused by reversing the direction where the particles collide on the wall.
The wave functions are given by three the types:
\begin{align}
\phi&=\cos k_1 x\\
\phi&=\sin k_2 x\\
\phi&=\cosh k_3 x,
\end{align}
where $k_1$, $k_2$, and $k_3$ satisfy following quantization conditions:
\begin{align}
-\frac{l}{L_\theta}&=k_1 l \tan \frac{k_1 l}{2}\label{eq:quantization_condition_of_NN1}\\
\frac{l}{L_\theta}&=k_2 l \cot \frac{k_2 l}{2}\label{eq:quantization_condition_of_NN2}\\
\frac{l}{L_\theta}&=k_3 l \tanh \frac{k_3 l}{2}\label{eq:quantization_condition_of_NN1}.
\end{align}
The approximation (\ref{eq:aproximation1}) and
\begin{align}
\tan x &= x - n \pi + O \big[ (x - n \pi ) ^3 \big]\label{eq:aproximation2}\\
\tanh x &= x  + O \big[ x ^3 \big]
\end{align}
lead to energy spectrum of each particle as
\begin{align}
E_n=\frac{ \kappa (n-1)^2}{l^2} - \frac{2\nu}{l},
\label{eq:Energy_of_N_N}
\end{align}
where $n \in \mathbb{N}$.
By using equation (\ref{eq:integral_approximation_of_integer}),
    we obtain partition function $Z^{NN}$ as follows:
\begin{align}
Z^{NN} &= \big\{ \sum_{n=1}^\infty \exp ( -\beta E_n) \big\}^N \notag\\
&= \exp \Big( \frac{2 N \nu \beta}{l}\Big)
    \Big\{ \int^\infty_0 \dd x \exp \big( -\frac{\beta\kappa}{l^2} x^2 \big) + \frac{1}{2}\Big\}^N \notag\\
&= \exp \Big( \frac{2 N \nu \beta}{l}\Big)
    \Big( \frac{l}{2}\sqrt{\frac{\pi}{\kappa \beta}} + \frac{1}{2} \Big)^N.
\end{align}
Let force acting on the wall be $p^{NN}$.
An equation of state in this case is given by
\begin{align}
\Big(
    p^{NN} + \frac{2 N \nu}{l^2} 
\Big)
\bigg(
    l 
    + \sqrt{\frac{\beta \kappa}{\pi}}
\bigg)
= \frac{N}{\beta}.
\label{eq:classical_equation_of_state_at_N-N}
\end{align}
We call the term $\frac{2 N \nu}{l^2}$ a force correction term and
the term $\sqrt{\frac{\beta \kappa}{\pi}}$ a length correction term.
Note that $\nu = 0$ implies the Neumann condition.
Let $p^{N}$ be force under the pair of the Neumann conditions,
    the equation of state is given by
\begin{align}
    p^{N}
\bigg(
    l 
    + \sqrt{\frac{\beta \kappa}{\pi}}
\bigg)
= \frac{N}{\beta}.
\label{eq:classical_equation_of_state_at_2N}
\end{align}
The above equation is different from
        the equation of the pair of the Dirichlet conditions (\ref{eq:classical_equation_of_state_at_D-D}).
\subsection{Three-dimensional box with a quasi-Neumann condition}
\label{sec:Three-dimensional box with a quasi-Neumann condition}
Here, we study a three-dimensional box with a quasi-Neumann condition at the wall of box.
Let $(x,y,z)$ be a three-dimensional Cartesian coordinate.
Consider a cuboid box whose median point is the origin of the coordinate.
Let its sides be parallel to the coordinate axes.
We introduce $l_x,l_y$, and $l_z$ as the lengths of the sides.
As is the previous section,
    we impose identical boundary conditions at the walls of the box: 
        the boundary condition at
            $x=\frac{l_x}{2}$,
            $y=\frac{l_y}{2}$,
            and
            $z=\frac{l_z}{2}$
        are characterized by $L_\theta$
        and 
        the boundary condition at
            $x=-\frac{l_x}{2}$,
            $y=-\frac{l_y}{2}$,
            and
            $z=-\frac{l_z}{2}$
        are characterized by $-L_\theta$.
We introduce quantum numbers for $x$, $y$, and $z$ directions as $n_x$, $n_y$, $n_z \in \mathbb{N}$,
which characterize energy of a particle, $E_{n_x,n_y,n_z}$, as
\begin{align}
E_{n_x,n_y,n_z} 
= \kappa 
    \bigg\{
        \frac{(n_x-1)^2}{l_x^2} + \frac{(n_y-1)^2}{l_y^2} + \frac{(n_z-1)^2}{l_z^2}
    \bigg\}
    - 2\nu 
    \bigg(
         \frac{1}{l_x} + \frac{1}{l_y} + \frac{1}{l_z}
    \Big).
\end{align}
The energy implies that force is different by the directions.
We introduce $p_x$ as force of the $x$ direction,
the equation of state is given by
\begin{align}
\bigg(
    p_x + \frac{2 N \nu S_x}{V^2}
\bigg)
\bigg(
    V + S_x \sqrt{\frac{\kappa\beta}{\pi}}
\bigg)
= \frac{N}{\beta},
\label{eq:classical_equation_of_state_at_3D}
\end{align}
where $S_x = l_y l_z$ and $V = l_x S_x$.
The force of the other directions is obtained by the same calculation.
Note that the terms showing the aeolotropic behavior of pressure vanish in the thermodynamic limit $V\to\infty$
        under the constant density condition, $\frac{N}{V}=$\ constant.
    The equation of state (\ref{eq:classical_equation_of_state_at_3D}) thus predicts no aeolotropic bulk gas
        but shows the dependence of boundary effects on the scale of the system.
\section{Conclusion and Discussion}          
We studied free particles in a closed box
    whose boundary conditions are generalized
        by self-adjoint extensions of the Hamiltonian.
We solved the energy spectrum conditions approximately
    and obtained the equations of state from the energy spectra by integral approximation.
            Note that our results are significant primarily for finite size systems since the  
boundary effects disappear in the thermodynamic limit; therefore, our  
equations of state do not describe the bulk properties of free  
particles at the thermodynamic limit.
\par
We assume three different combinations of the boundary conditions.
First, we considered the one-dimensional box
    with the Dirichlet condition and a quasi-Neumann condition.
We assumed three types of statistics: Maxwell-Boltzmann, Bose-Einstein, and Fermi-Dirac statistics,
    and obtained the equations of state
        (\ref{eq:classical_equation_of_state_at_N-D}) and (\ref{eq:quantum_equation_of_state_at_D-N})
    and evaluated 
        the behavior of the equations in the limit $l \to 0$,
            arriving at
                (\ref{eq:lim_of_classical_equation_of_state}),
                (\ref{eq:lim_of_bose_equation_of_state}),
                and
                (\ref{eq:lim_of_fermi_equation_of_state}).
Although these five equations are different in
        statistics 
        and
        in the region of $l$,
            they share the same force correction terms.
The common force correction terms
have the same dependence on length as that of van der Waals equation.
The force correction terms are dependent on the boundary parameter,
    whereas other terms are independent.
\par
Second,
    we studied Maxwell-Boltzmann particles under the Dirichlet condition at both of the walls. 
We found that
        the equation of state (\ref{eq:classical_equation_of_state_at_D-D}) is
                no longer that of ideal gas 
due to the length correction term in (\ref{eq:classical_equation_of_state_at_D-D}),
        which is different from that of van der Waals equation of state (\ref{eq:vdW_equation_of_state}).
        It has been studied in \cite{Tsutsui2,Tsutsui3} that
    there is a difference between two pressures, one of which obtained under the pair of the Dirichlet conditions and 
    the other evaluated under the pair of the Dirichlet and the Neumann conditions.
The equations of state of
    (\ref{eq:classical_equation_of_state_at_N-D})
    and 
    (\ref{eq:classical_equation_of_state_at_D-D}) 
        lead to the high temperature limit of the pressure difference (\ref{eq:tsutsui_pressure}),
            which is an increasing function of temperature and vanishes in the thermodynamic limit.
\par
Third,
    we studied particles obeying Maxwell-Boltzmann statistics
        under a quasi Neumann condition at both of the walls.
We obtained the equation of state (\ref{eq:classical_equation_of_state_at_N-N}),
        which has both the length correction term and the force correction term.
The strength of the force correction term
    in this case is twice of that of the pair of the Dirichlet condition and a quasi Neumann condition.
The length correction term in this case
is similar to that of the pair of the Dirichlet conditions except for its sign. 
Note that the equation of the pair of the Neumann conditions (\ref{eq:classical_equation_of_state_at_2N})
            is different from that of the pair of the Dirichlet conditions.             
\par
These three results clearly show
that boundary conditions can yield
    force correction terms that have inverse square dependence
    on the length, analogously to van der Waals equation.    
However, the force correction term in van der Waals equation and those of our equations are different
    in particle number dependence.
Our equations of state are influenced
    by boundary effects which vanish in the thermodynamic limit,
        and hence they are meaningful only in the finite length system.    
In the case of van der Waals equation,
    the order $N^2$ is regarded as the number of the inter-molecular interactions,
    which is $N(N-1)\sim N^2$ since the $N$ particles interact with each other.
On the other hand,
    the force correction term of the pair of the Dirichlet condition and a quasi Neumann condition
        is in proportion to $N$
            and 
    that of the pair of quasi Neumann conditions
        is twice of that of the pair of the Dirichlet condition and a quasi Neumann condition.
There is no force correction term in the equation of state of the pair of the Dirichlet conditions.
Interestingly, the product of the particle number and the number of the walls of a quasi Neumann condition equals
    the dependence of the particle number of the force correction terms in our model.
This implies that
    the particle number dependence can be interpreted as the number of the interactions, 
            as in the case with van der Waals equation.
\par
We also briefly touched upon
    the three-dimensional box with a quasi-Neumann condition.
The obtained equation of state (\ref{eq:classical_equation_of_state_at_3D})
    is a three-dimensional extension of
        the one-dimensional equation (\ref{eq:classical_equation_of_state_at_N-N})
    and we find that
        the three-dimensional equation possesses
            the same property of the one-dimensional equation,
    confirming that boundary effects can appear in three dimensions, too.
Finally, we mention that generalized
    boundary conditions have been used
        for effective descriptions of rapidly varying potentials at small scales,
which are approximately represented by the vanishing limit of widths of step functions \cite{Tsutsui6,Cheon1,Cheon2}.
There, the boundary condition (2.7) and parameter $L_\theta$ are derived
    from such limits of two step functions \cite{Tsutsui6}.
            These circumstances may appear, for instance,
    at junctions of two semiconductor layers
    and, if so, our boundary effects may actually be observed experimentally.   
\section{Acknowledgements}
The author would like to thank Prof. I. Tsutsui and Dr. Y. Abe, Dr. T. Ichikawa, T. Sasaki and H. Imazato
for useful comments.
\appendix
\section{On approximation}
\label{sec:approximation}
In this appendix,
    we give a brief account of integral approximation of the sum of series.
For all $n \in \mathbb{Z}$,
    $f(x)$ can be expanded as
\begin{align}
f(x) 
= \sum_{k=0}^\infty 
    \frac{1}{ k !}
    \Big( x - n - \frac{1}{2} \Big)^k f^{(k)}(n + \frac{1}{2}),
\end{align}
where $ x \in (n,n+1)$.
We obtain
\begin{align}
\int_0^\infty f(x) dx
= \sum_{k=0}^\infty
    \frac{1}{(2k + 1)!}
    \frac{1}{2^{2k}}
    T_{2k},
\end{align}
where 
\begin{align}
T_{k}:= \sum _{n=0} ^\infty f^{(k)}(n+\frac{1}{2}).
\end{align}
We apply the above equation to $f^{(2k)}(x)$
and eliminate the differential term by successive iteration,
then we obtain the following series expansion:
\begin{align}
\int_0^\infty f(x) dx 
=
    \sum_{n=1}^\infty a_{2n-1} f^{(2n-1)}(0) 
    + \sum_{n=0}^\infty f(n+\frac{1}{2}),
\label{eq:integral_approximation_of_one_half}
\end{align}
where
\begin{align}
\kappa_{k}:=&
    \frac{1}{k!}\frac{1}{2^k}\notag\\
a_{2k-1} =&
    2 \kappa_{2k+1}
    - \sum_{n=1}^{k-1}
        2 \kappa_{2n+1}
        a_{2(k-n)-1}.
\end{align}
The coefficient $a_{2k-1}$ tends to $0$ rapidly in the limit $k \to \infty$.
In addition, $f(x) = F(\sqrt{\alpha} x)$ in our system;
therefore
\begin{align}
f^{(n)}(0) \sim O[\alpha^{\frac{n}{2}}].
\end{align}
From the above equations,
    then we can
    neglect
    the sum of the first term of the right hand side of
    (\ref{eq:integral_approximation_of_one_half}) if $\alpha <  1$.
\par
Now, we show a similar formula to (\ref{eq:integral_approximation_of_one_half}).
Let f(x) expand as follows:
\begin{align}
f(x) 
= \sum_{k=0}^\infty 
    \frac{1}{ k !}
    \Big( x - n \Big)^k f^{(k)}(n).
\end{align}
We define $S_k := \sum _{n=0} ^\infty f^{(k)}(n)$ and repeat the above discussion,
then the following equation is obtained:
\begin{align}
\int_0^\infty f(x) dx
&=
    \sum_{k=0}^\infty 
        \frac{1}{(k+1)!} \frac{1}{2^{k+1}}f^{(k)}(0) 
    +
    \sum_{k=1}^\infty
        \frac{1}{(2k + 1)!}
            \frac{1}{2^{2k}} \sum _{n=0} ^\infty f^{(2k)}(n)\notag\\
&=
    \sum_{n=1}^\infty
        f(n)
    +\sum_{n=0}^\infty
      b_n f^{(n)}(0),
    \label{eq:integral_approximation_of_integer}
\end{align}
where
\begin{align}
b_{0} & = \frac{1}{2}\notag\\
b_{2k} &= \frac{1}{2} a_{2k-1}\notag\\
b_{2k-1} &= a_{2k-1} + \kappa_{2k} - \sum_{n=1}^{k-1} \kappa_{2n} a_{2(k-n)-1}.
\end{align}
\section{On polylogarithm}
\label{sec:polylog}
The polylogarithm\cite{Wood}, $\textrm{Li}_s (z)$, is defined in $|z|<1$ as follows:
\begin{align}
\textrm{Li}_s (z) = \sum_{k=1}^\infty \frac{z^k}{k^s}.
\end{align}
In $|z|>1$,
it is defined by
    analytical continuation
of the above equation.
It satisfies the following formulae:
\begin{align}
\textrm{Li}_s (-z) + \textrm{Li}_s (z) &= 2^{1-s} \textrm{Li}_s (z^2)
\\
\lim_{\textrm{Re}(\mu)\to\infty} \textrm{Li}_s (e^\mu) &= - \frac{\mu^s}{\Gamma(s+1)};
\end{align}
therefore we obtain the asymptotic behavior of them:
\begin{align}
\textrm{Li}_{\frac{1}{2}} \big\{-\exp (z) \big\} &\to 
- 2 \sqrt{\frac{z}{\pi}}\\
\textrm{Li}_{\frac{3}{2}}\big\{-\exp \big( z \big) \big\} &\to 
- \frac{4 z }{3} \sqrt{\frac{z}{\pi}}.
\end{align}
We can express $\textrm{Li}_{\frac{1}{2}}(\pm e^y )$ and $\textrm{Li}_{\frac{3}{2}}(\pm e^y )$ in integral forms as
\begin{align}
\textrm{Li}_{\frac{1}{2}} (\pm e^y ) &= 
\int^\infty_0 \frac{2 \ \dd x}{\pm\exp[\pi x^2-y]-1}\label{eq:integral_form_of_polylog_0.5}\\
\textrm{Li}_{\frac{3}{2}} (\pm e^y ) &= 
\int^\infty_0 \frac{4 \pi x^2 \ \dd x}{\pm\exp[\pi x^2-y]-1}\label{eq:integral_form_of_polylog_1.5}.
\end{align}
The above equations imply that
    $\textrm{Li}_{\frac{1}{2}}(\pm e^y )$ and $\textrm{Li}_{\frac{3}{2}}(\pm e^y )$ are monotone functions
        because of the integrands are monotone functions of $y$.
\newpage

\end{document}